\documentclass[runningheads]{svmult}
\usepackage{makeidx}   
\usepackage{graphicx}  
\usepackage{subeqnar}  
\usepackage{multicol}  
\usepackage{cropmark} 
\usepackage{physprbb}  
\def\lesssim{\mathrel{\hbox{\rlap{\hbox{\lower4pt\hbox{$\sim$}}}\hbox{$<$}}}}
\def\gtrsim{\mathrel{\hbox{\rlap{\hbox{\lower4pt\hbox{$\sim$}}}\hbox{$>$}}}}

\newcommand{\ffrac}[2]
  {\left( \frac{#1}{#2} \right)}

\begin{document}
\title*{ Detecting Ultra High Energy Neutrinos \protect\newline by  Upward Tau Airshowers and Gamma Flashes }
\toctitle{Detecting Ultra High Energy Neutrinos:\newline by Upward
Tau and Gamma Flashes Airshowers }
%
%
\titlerunning{Tau Airshowers by UHE neutrinos}
%
\author{Daniele Fargion\inst{1}
}
\authorrunning{Daniele Fargion}
%
%
\institute{Physics Department and INFN , Rome University 1,\\
     Pl.A.Moro 2, 00185, Rome, Italy}

\maketitle              


\begin{abstract}
 $\tau$ Air-showers are the best trace of rarest Ultra High Energy neutrinos UHE
$\nu_{\tau}$, $ \bar\nu_{\tau} $ and $ \bar{\nu}_e $ at PeV and
higher energy.   $\tau$ Air-showers may generate  billion times
amplified signals  by their secondaries .   Horizontal amplified
$\tau$ air-showers by $\nu_{\tau} N$ and  UHE  $\bar\nu_{e} e$  at
PeV emerging from mountain chain might be the most power-full
imprint. Upward UHE $ \nu_{\tau} N $ interaction on Earth crust at
horizontal edge and from below, their consequent  UHE $\tau$
air-showers beaming toward high mountains
 should  flash $\gamma$,$\mu$,X and optical detectors on the top.
  Upward $\tau$ air-shower may hit   nearby satellite  flashing them by short,
  hard, diluted $\gamma-$burst at the edge of
 Gamma Ray Observatory BATSE   threshold.
We identify these events with recent (1994) discovered upward
Terrestrial Gamma Flashes  (TGF) and we probed their UHE $\tau$ -
UHE $\nu_\tau$ origin. From these TGF data approximated UHE
$\nu_\tau$ flux and $\Delta m_{\nu_\mu \nu_\tau}$ sever lower
bound are derived. Partial TGF Galactic signature is also manifest
within known 47 TGF events at ${ \simeq 2\cdot
10^{-3}}$probability. Well known $X-\gamma-\mathrm{TeV}$ active
galactic and extragalactic sources have found probable counterpart
in  TGF arrival directions. Detection of elusive UHE $\nu_\tau$
seem finally achieved.

\end{abstract}

\section{The UHE $\bar{\nu}_e$, $\nu_\tau$, $\bar{\nu}_\tau$ interactions and their UHE $\tau$
secondary}
Ultra high energy astrophysical neutrino (UHE$\nu$) from PeVs
($\gtrsim 10^{15}$ eV) up to  ($10^{18}$ eV) EeV and GZK cut off
energies ($ \gtrsim 10^{19}$ eV) might be traced  by $\tau$
induced air showers and by their millions to hundred billions
multiplicity in secondaries particles, ($N_{\mu} \sim 10^6 \left(
\frac{E_{\nu}}{PeV} \right)$, $N_{X} \sim 10^{10} \left(
\frac{E_{\nu}}{PeV} \right)$, $N_{opt} \gtrsim 10^{11} \left(
\frac{E_{\nu}}{PeV} \right)$).\\ Indeed astrophysical PeVs UHE
anti-neutrino electrons, $\bar{\nu}_e$, near the Glashow W
resonance peak, $E_{\bar{\nu_e}} = M^2_W / 2m_e \simeq 6.3 \cdot
10^{15}\, eV$, (dominant over expected UHE PeV atmospheric
neutrino signals), may be observable by their secondary horizontal
$\tau$ air showers originated by UHE chain reaction $\bar{\nu}_e +
e \rightarrow W^- \rightarrow \bar{\nu}_{\tau} + \tau^-$ inside
the concrete rock of a high mountain. Also UHE  $\nu_\tau$,
$\bar\nu_{\tau} $ at ($10^{16}$ - $10^{17}$eV) interacting with
nuclear matter ($\nu_\tau$ $N$) must be observable if flavor
mixing $\nu_{\mu}\leftrightarrow \nu_{\tau}$ take place as shown
by Superkamiokande data, because of huge astrophysical distances
respect to oscillation ones.\\
 Therefore UHE
 $\nu_{\tau}$ and $\bar{\nu_{\tau}}$ may be converted and they may
 reach us from high energy galactic sources, as pulsars, SNRs,
 black holes or galactic SGRs microquasars, as well as from powerful
 extragalactic AGNs, quasars or GRBs, even for any small mass
 mixing, ($\Delta m_{ij}^2 \sim 10^{-4}$ eV $^2$) or any high (GZK)
 energy because of the large galactic (Kpcs) and extreme cosmic
 (Mpcs) distances:

 \begin{equation}
 L_{\nu_{\mu} - \nu_{\tau}} = 4 \cdot 10^{-3} \,pc \left(
 \frac{E_{\nu}}{10^{16}\,eV} \right) \cdot \left( \frac{\Delta m_{ij}^2
 }{(10^{-2} \,eV)^2} \right)^{-1}
 \end{equation}

 Rare upward UHE $\tau$, born by $\nu_{\tau}$ and
$\bar{\nu_{\tau}}$ nuclear (or rare $\bar{\nu_e} - e$ interactions
near the upward earth surface), may escape outside on air where
they may spontaneously decay triggering upward vertical, oblique
or near horizontal $\tau$ air showers. The vertical ones (by small
 nadir angle) occur preferentially at low energies nearly
 transparent to the Earth ($E_{\nu} \sim 10^{15} - 10^{16} $ eV). The
 oblique $\tau$ air showers whose arrival directions have large
 nadir angle, are related mainly to higher energy $\nu_{\tau}$, or
 $\bar{\nu_{\tau}}$ nuclear interactions ($E_{\bar{\nu_{\tau}}}
 \geq 10^{17} - 10^{20}$ eV). Indeed these horizontal - upward UHE
 $\nu_{\tau}$ cross a smaller fraction of the Earth volume and
 consequently they suffer less absorption toward the horizon. \\
These huge horizontal or upward air-shower signals being at least
million to billion times more abundant than the original and
unique UHE $\tau$  or UHE $\mu$ track in underground Km cube
detectors are much easier to be discovered with no ambiguity. We
remind that long tracks in km$^3$ detectors are mostly noisy
signals by TeVs to tens of TeVs muons secondaries generated by
atmospheric neutrinos born by common cosmic ray interactions in
upper atmosphere. The $\tau$ air shower is analogous to the
Learned and Pakwasa $(1995)$ "double bang" in underground neutrino
detectors. The novelty of the present "one bang in" (the rock) -
"one bang out" (the air) lays in the self-triggered explosive
nature of $\tau$ decay in flight and its consequent huge amplified
air shower signal  at a characteristic few Kms distance.
Detectable horizontal gamma bursts (mainly bremsstrahlung photons)
are among the most abundant signal.
 The source of UHE $\nu_{\tau}$, since Super Kamiokande evidence of
neutrino flavour mixing, must be as abundant as muon $ \nu_{\mu}$
ones. \\ Moreover the expected $ \nu_{\tau} $ signals, by their
secondary tau tracks at highest cosmic ray energy window $1.7\cdot
10^{21} \,eV > E_{\tau} > 1.6\cdot 10^{17} \, eV $, must exceed
the corresponding $ \nu_{\mu} $ (or muonic) ones, making UHE $
\nu_{\tau} $ above $0.1$ EeV the most probable UHE signal. Indeed,
the Lorentz-boosted tau range length grows (linearly) above muon
range, for $ E_{\tau} \geq 1.6 \cdot 10^8 GeV $; (see Fig (1)
eq.3): the tau track reaches its maxima extension, bounded not by
pair production (eq. 2), but by growing nuclear electro-weak
interactions (eq. 4), $ R_{\tau_{\max}} \simeq 191\;Km$, at energy
$ E_{\tau} \simeq 3.8\cdot 10^9\;GeV$ .

\begin{equation}\label{6}
 R_{R_{\tau}} \cong 1033 \; Km
\left(\frac{\rho_r}{5}\right)^{-1} \cdot
 \left\{\, 1 \,+\, \frac{\ln\left[\left(\frac{E_{\tau}}{10^8 \,
\mathrm{GeV}}\right)\left(\frac{E_{\tau}^{\min}}{10^4 \,
\mathrm{GeV}}\right)^{-1}\right]}{(\ln \, 10^4 )}\right\} .
\end{equation}



\begin{equation}\label{7}
R_{\tau_o} = c \tau_{\tau} \gamma_{\tau} = 5 \, \mathrm{Km} \,
\left(\frac{E_{\tau}}{10^8 \, \mathrm{GeV}}\right) \; .
\end{equation}

\begin{equation}\label{9}
 R_{W_{\tau}} = \frac{1}{\sigma N_A \rho_r} \simeq
\frac{2.6\cdot 10^3 \, \mathrm{Km}}{\rho_r} \,
\left(\frac{E_{\tau}}{10^8\, \mathrm{GeV}}\right)^{-0.363} \; .
\end{equation}

 At this peak the tau range is nearly $20$
times longer than the corresponding muon range (at the same
energy) implying, for comparable fluxes, a ratio 20 times larger
in $ \nu_{\tau} $ over $ \nu_{\mu} $ detection probability. This
dominance, may lead to  a few rare spectacular event a year (if
flavor mixing occurs) preferentially in horizontal plane in
underground $Km^3$ detectors. The Earth opacity at those UHE
regimes at large nadir angles (nearly horizontal, few degree
upward direction) is exponentially different for UHE muons respect
to tau  at GZK  energies (corresponding to 500 Kms UHE Tau
lenghts), making the muon/tau flux ratio of such lenghts severely
(half billion time) suppressed.

 UHE $\nu$ above GZK are transparent to BBR cosmic
photons and they may easily reach us from far cosmological
distances. Therefore the puzzle of UHECR above GZK cut off may be
solved assuming that neutrinos (possibly of heaviest Muon/Tau
nature)share a light mass of few eV , in the frame-work of Hot
Dark Matter halos clustered around galaxies. Such light neutrinos
may form a huge hidden dark calorimeter  able to beam dump UHE
$\nu$ via $Z$ (s-channel), via  virtual $W$ ($t$ channel)  or W
pair productions. The corresponding cross sections for such $\nu$
$\nu$ interactions are shown in Fig. 2; their secondaries may be
final UHE anti-protons (or anti-neutrons) or UHE protons (or
neutrons) (Fargion,Mele,Salis 1997-1999) responsible of final
observed UHECR above GZK cut off. The interaction efficency by
relic light neutrinos via UHE $\nu$ at GZK cut off is thousands
times larger than UHE $\nu$ interactions on Earth atmosphere
and/or direct UHECR (nucleons,nuclei) propagations above GZK
distances. Therefore light neutrino mass may explain both hot dark
matter and UHECR above GZK (as well as their recent clustering in
triplets or doublets). Just to emphasize the $\nu$ mass roles in
high energy astrophysics, we remind the important case of a SN
MeVs neutrino burst arriving slowed by its mass relativistic
flight and its delayed arrival from far SN (galactic or better
extragalactic) events respect to the massless (prompt coeval)
gravitational waves. The expected time delay between the massless
graviton wave burst (by supernova quadrupole emission at distance
L and the $\nu_e$ neutronization neutrino burst), will be an
additional tests test to the elusive mass detection: $ \Delta t
\sim 50$ sec $ \ffrac{E_{\nu}}{5\,MeV}^{-2}
\ffrac{m_{\nu}}{5\,eV}^{2} \ffrac{L}{Mpc}$. (Fargion 1981). Let us
remind that massive neutrino imply new right handed interactions
in early Universe (Antonelli, R.Konoplich, D.Fargion 1981) and
multifluid gravitational clustering during galaxy formation epochs
(D.Fargion 1981, 1983).

UHE Tau $ E_{\tau} \geq \ 10^5 GeV - 5 \cdot 10^7 GeV  $
air-shower in front of high mountains chains will be easily induce
peculiar horizontal UHE $\tau$ decay beyond a thick mountain
(Fargion, Aiello, Conversano 1999). The high mountain act as a
clever filter:
\\a) as a wide angle screen of undesirable horizontal Ultra High Energy
Cosmic Rays (UHECR) (electro-magnetic shower, secondary Cherenkov
photons and muons),b) as a calorimeter for UHE $\nu_\tau$,$
\bar\nu_{\tau} $ and $ \bar{\nu}_e $, c) as a distance meter
correlating tau relativistic track and birth and its air-shower
opening  distance from the mountain with UHE tau original energy.
 d) as an unique source, by tau electromagnetic showering, of
horizontal rich, sharp, $\mu$second burst $\gamma$, X flash,
electron pairs and Cherenkov showers source. e) as an unique
source, by tau hadronic showering, of additional horizontal dense
muon pairs sharp bundle burst.
\\ An hybrid detector (gamma/optical)would get precise signal and arrival direction.
Because of the different neutrino interactions with energy and
flavors it will be possible to estimate, by stereoscopic,
directional and time structure signature, the spatial air-shower
origination in air, the primary tau distance decay from the
mountain (tens of meter for PeVs UHE $ \bar{\nu}_e $ and nearly
hundred meters up to  Kms for UHE $\nu_\tau$,$ \bar\nu_{\tau} $) $
E_{\tau} \geq \ 10^5 GeV , 5 \cdot 10^7 GeV $ , the consequent
most probable original UHE tau range and energy. Additional energy
calibration may be derived sampling shower intensities.\\ Hundreds
of detectors in deep wide valley would be necessary to get tens
taus of event a year.
\\ Screening by undesirable lateral
or downward noisy cosmic rays or natural radiation is possible by
directional and time clustering filter; therefore the induced
$\bar{\nu}_e e \rightarrow \tau$ air shower even in absence of
$\nu_{\mu} \leftrightarrow \nu_{\tau}$ oscillation should be
identified and detectable soon. Its unique $\bar{\nu}_e$ origin is
marked by the peaked  W resonance, and by the small mountain
$\bar{\nu}_e$ opacity and its high neutrino cross-section. Its
identity is marked by the expected fine tuned PeV energy at W peak
and the tau air-shower birth place near (a hundred meter) the
mountain wall.\\ More copious ($> 5$ times more) events by PeV up
to tens PeV $\;$ $\nu_{\tau} N$ interaction occur if, as most of
us believe, $\nu_{\mu}$ oscillate in $\nu_{\tau}$.\\
 It will be also possible to discover UHE $\tau$,
by observing the upward tau air-shower arriving from hundred
Kilometers away (near horizontal edges) from high mountains, high
balloon and satellites; such UHE tau created within a wide (tens
thousands to millions square km$^2$ wide and hundred meter UHE Tau
depth in Earth crust) target would discover easely UHE
$\nu_\tau$,$ \bar\nu_{\tau} $ neutrinos at PeV up to EeV energies
and above, just within the mysterious GZK frontiers. The discover
will need capable gamma, optical and mainly muon bundle detectors
within present technology as studied elsewhere.\\ From the same
highest mountains, balloons and near orbit satellite, looking more
downward toward the Earth  it is possible to discover more
frequent but lower energetic astrophysical $\simeq$ PeV - tens PeV
 neutrinos still nearly transparent to the Earth volumes and
(Gandhi et al. 1998),(see Fig.3). \\ The UHE neutrinos
$\bar{\nu_e}$,${\nu}_{\mu}$ $\bar{\nu}_{\mu}$ are default and
expected UHECR ( $\gtrsim 10^{16}$ eV) secondary products near AGN
or microquasars by common photo-pion decay relics by optical
photons nearby the source (PSRs, AGNs) ($p + \gamma \rightarrow n
+ \pi^+, \pi^+ \rightarrow \mu^+ \nu_{\mu}, \mu^+ \rightarrow e^+
\nu_e \bar{\nu}_{\mu} $), or by proton proton scattering in
galactic interstellar matter. The maximal observational distances
from mountains, baloons or satellites, may
 reach $\sim$ 110 Km $(h/Km)^{\frac{1}{2}}$ toward the horizon,
 corresponding to a UHE $\tau$ energy $\sim 2 \cdot 10^{18}$ eV $(h/ Km)^\frac{1}{2}$.
 Therefore we propose to consider such upward shower nearly horizontal detection
 from high mountains to test this highest $\nu_{\tau}
 \bar{\nu_{\tau}}$ energy window almost opaque to Glashow UHE
 $\bar{\nu_e}$ fluxes.

\section{Upward $\tau$ air shower in Terrestrial Gamma Flash}

The expected downward muon number of events $ N_{ev} (\bar{\nu}_e
e\to\bar{\nu}_{\mu} \mu)$ in the resonant energy range, in Km$^3$,
[Table 7,The Gandhi et all,1998] 
was found to be $N_{ev} = 6$ a year. One expect a comparable
number of reactions $ (\bar{\nu}_e e\to \bar{\nu}_{\tau} \tau) $.
However the possible presence of primordial $\nu_{\tau}, \,
\bar{\nu_{\tau}}$ by flavor mixing and $\nu_{\tau}, \,
\bar{\nu_{\tau}} N$ charged current interactions lead to a factor
5 larger rate, $ N_{ev} = 29$ event/year.
\\ If one immagines a gamma/optical detector at 5 km far in front
of the Alps Argentier mountains (size 10 km, height 1 km) one
finds a $\tau$ air shower volume observable within a narrow
 beamed cone (Moliere radius $\sim 80$ m / distance $\sim 5$ Km):
 ($\Delta \theta \sim
 1^o$, $\Delta \Omega \sim 2 \cdot 10^{-5}$) and
 an effective volume $V_{eff} \simeq 9 \cdot 10^{-5} $
 Km$^3$ for each observational detector (Fargion 2000).
  Each one is comparable to
 roughly twice a Super Kamiokande detector.
  We expect, following AGN - SS91 model [The Gandhi et all,1998] 
  a total rate of: (6) ($\bar{{\nu}_e} e$) + (29) ($\nu_{\tau} N$) = 35 UHE $\nu_{\tau}$ event/year/Km$^3$;
 at energies above 3 PeV we may expect a total rate of N$_{ev}$ $\sim$
 158 event/year in this Alps Argentiere mountains valley and
 nearly 3.2 $\cdot 10^{-3}$ event/year for each detector.
 In a first approximation, neglecting Earth opacity, it is possible to show that the Earth volume observable from the top of
a mountain at height $h$, due to UHE $\tau$ at 3 PeV crossing from
below, is approximately V $\approx 5 \cdot 10^4$ Km$^3$
$\ffrac{h}{Km} \ffrac{E_{\tau}}{3\,PeV}$. The upward shower would
hit the top of the mountain. For the same $\tau$ air shower
beaming ($\Delta \theta \sim 1^o$, $\Delta \Omega \sim 2 \cdot
10^{-5}$) we derive now an effective volume $\sim$ 1 Km$^3$.
Therefore a detector open at $2 \pi$ angle on a top of a 2 Km
height mountain may observe nearly an  event every two month from
below the Earth. The gamma signal above few MeV would be
(depending on arrival nadir angle) between  $3\cdot 10^{-2}$
cm$^{-2}$ (for small nadir angle) to $10^{-5}$ cm$^{-2}$ at far
distance within 3 PeV energies. A contemporaneous (microsecond)
optical flash ($\gtrsim 300 \div 0.1 \, cm^{-2}$) must occur.
Keeping care of the Earth opacity, at large nadir angle ($\gtrsim
{60}^0$) where an average Earth density may be assumed ($< \rho >
\sim 5$) the transmission probability and creation of upward UHE
$\tau$ is approximately

\begin{equation}
P(\theta,\, E_{\nu}) = e^{\frac{-2R_{Earth} \cos
\theta}{R_{\nu_{\tau}}(E_{\nu})}} (1 - e^{-
\frac{R_{\tau}(E_{\tau})}{R_{\nu_{\tau}}(E_{\nu})}}) \, .
\end{equation}

This value, at PeV is a fraction of a million.\\ The corresponding
angular integral effective volume observable from a high mountain
(or baloon) at height $h$ (assuming a final target terrestrial
density $\rho = 3$)  is:

\begin{footnotesize}
\begin{equation}
  V_{eff} \approx 0.3 \, Km^3 \ffrac{\rho}{3}\ffrac{h}{Km} e^{-
  \ffrac{E}{3\,PeV}}
  \ffrac{E}{3\,PeV}^{1.363}
\end{equation}
\end{footnotesize}

This rate is quite large and the expected $\tau$ air airshower
signal (gamma burst at energies $\gtrsim 10 \, MeV$) should be
$\phi_{\gamma} \simeq 10^{-4} \div 10^{-5}$ cm$^{-2}$, while the
gamma flux at ($\sim 10^5 \, eV$) or lower energies (from electron
pair bremsstrahlung) may be two order of magnitude larger.\\ The
optical Cherenkov flux is large $\Phi_{opt} \approx 1$ cm$^{-2}$.
The tau upward air showers born in a narrow energy
 window, $10^{15}$ eV $ \lesssim E_{\nu} \lesssim  5 \cdot 10^{16} $
 eV (Fig.3) may penetrate high altitude leaving rare beamed upward gamma
 shower bursts whose sharp ($\sim $ hundreds $\mu$sec because of the hundred kms high
 altitude shower distances) time
 structure and whose hard ($\gtrsim 10^{5}$eV) spectra may hit near
 terrestrial satellites.
We claim (Fargion 2000) that such gamma upward events originated
by tau air
 showers produce gamma bursts at the edge of GRO-BATSE sensitivity
 threshold. In particular we argue that very probably such upward
 gamma events have been already detected since April 1991 as
 serendipitious sharp ($\lesssim 10^{-3}$ sec) and hard ($\gtrsim
 10^5$ eV) BATSE gamma triggers originated from the Earth and
 named consequently as Terrestrial Gamma Flashes (TGF).
The visible Earth surface from a satellite, like BATSE, at height
$h \sim 400$ Km and the consequent effective volume for UHE
$\nu_{\tau} N$ PeVs interaction and $\tau$ air shower beamed
within $\Delta \Omega \sim 2 \cdot 10^{-5} rad^2$ is: (note
$<\rho> \simeq 1.6$ because 70 \% of the Earth is covered by seas)
 $ V_{eff} = V_{TOT} \Delta \Omega \simeq 60 \, Km^3.$
The effective volume and the event rate should be reduced, at
large nadir angle ($\theta > 60^o$) by the atmosphere depth and
opacity (for a given $E_{\tau}$ energy). Therefore the observable
volume may be reduced approximately to within 15 Km$^3$ values and
the expected UHE PeV event rate is
\begin{equation}
  N_{ev} \sim 150 \cdot e^{-
  \ffrac{E_\tau}{3\,PeV}}
  \ffrac{E_\tau}{3\,PeV}^{1.363}
  \ffrac{h}{400 Km}\qquad
  \frac{\mathrm{events}}{\mathrm{year}}
\end{equation}

The TGF signals would be mainly $\gamma$ at flux $10^{-2}$
cm$^{-2}$ at X hundred keV energies. The observed TGF rate is
lower by nearly an order of magnitude, and this suggests higher
$E_{\nu}$ energies (to overcome BATSE threshold) and consequently
small additional probability suppression leading to observed TGF
events rate. However since
 1994 (Fishman et al.) TGF understanding of presently known 75
 records over nearly eight thousand BATSE triggers is based on an
 unexpected and mysterious high latitude lightening of geophysical
 nature (the so called "Sprites" or "Blue Jets"). We do not
 believe in that interpretation.
 We notice that among the 75 records only 47 are published in
 their details, while 28 TGF are still unpublished. Their data
 release is therefore urgent and critical.
 While Blue Jets might be in principle triggered by upward tau
 air showers in the atmosphere (a giant "Wilson" room) we believe they are
 not themselves source of TGF. In particular their observed
 characteristic propagation velocity ($\lesssim 100$ Km/s) from
 distances $\sim$ 500 Km, disagree with short TGF millisecond
 timing and would favor a characteristic TGF time of few seconds. \\
  Moreover TGF data strongly disfavor by its hard spectra
 the terrestrial Sprites connection. On the contrary the expected
 UHE tau upward air showers lead to a gamma burst flux,
 spectra, and fine time structure fluence in agreement with the
 observed TGF ones and in agreement with the expected
 flux models.
 The correlations of these clustered TGFs directions
 toward \\ (1) well known and maximal powerful galactic and
 extra-galactic sources either at TeV, GeV-MeV, X band ,(2) recent
 first anisotropy discovered on UHECR at EeV by AGASA, (see Fig.4, and
 Hayoshida et al. 1999) (3) the
 Milky Way Galactic Plane,  support
 and make compelling the TGF  identification as secondary gamma
 burst tail of UHE $\tau$ induced upward air shower.
   The present TGF-$\tau$ air-shower identification could not
    be produced by UHE $\bar{\nu}_e$ charged current
   resonant event at ($E_{\bar{\nu_{e}}} = M^2_W / 2m_e = 6.3 \cdot 10^{15}$
   eV), and therefore it stand for the UHE $\nu_{\tau}
 \bar{\nu_{\tau}}$ existence. Consequently it gives support to the
 Superkamiokande evidences for $\nu_{\mu}\leftrightarrow \nu_{\tau}$
  flavor mixing from far PSRs or AGNs sources toward the Earth.
     At the present the very probable $\nu_{\tau}
     \bar{\nu_{\tau}}$ source of TGFs and their probable partial galactic location
     infer a first lower bound on
     $\Delta_{m_{\nu_{\mu} \nu_{\tau}}}$ ($L < 4$ Kpc, $\Delta_{m_{\nu_{\mu} \nu_{\tau}}}
      > 10^{-8}$ eV$^2$)
     and it offers a first direct test of
     the same existence of the last missing, yet unobserved,
     fundamental neutral lepton  particle: $\nu_{\tau}$ and $ \bar{\nu_{\tau}}$ . \\









\begin{figure}[]
\begin{center}
 \includegraphics[width=0.8\textwidth, height=5cm, bb=173 144 625 533] {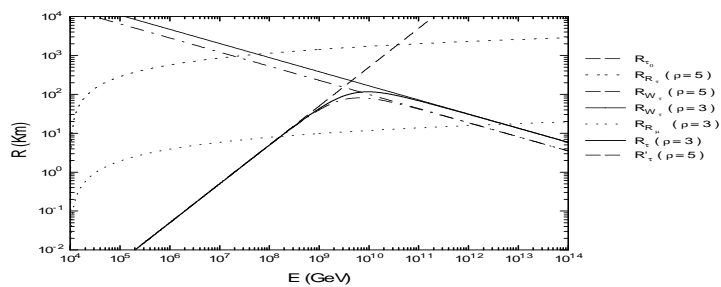}
\end{center}
  \caption{The tau ranges as a function of the tau energy respectively for
    tau lifetime (dashed line) $R_{\tau_o}$, for tau radiation
    range $R_{R_\tau}$ ,(short dashed line above) and tau
    electroweak interaction range $R_{W_\tau}$, for two densities
    $\rho_r$ (long dashed lines, continuous) and their combined range
    $R_{\tau}$. Below the corresponding radiation range
    $R_\mu$ for muons (dotted line).}
\label{fig:boxed_graphic 1}
\end{figure}

\begin{figure}[]
\begin{center}
\includegraphics[width=0.8\textwidth, height=4.5cm]{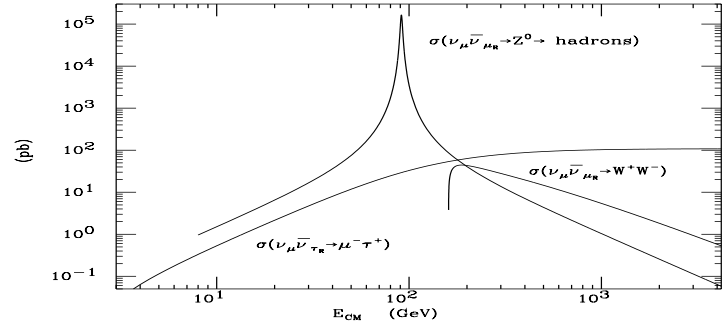}
\end{center}
  \caption{The total cross sections for the UHE $ \nu\bar{\nu}$ indicated processes
as function of the center of mass energy (for a relic neutrino
mass $m_\nu = 10 eV$)(Fargion,Mele,Salis 1999)}
\label{fig:boxed_graphic 2}
\end{figure}



\begin{figure}[]
\begin{center}
\includegraphics[width=0.8\textwidth , height=4.5cm]
{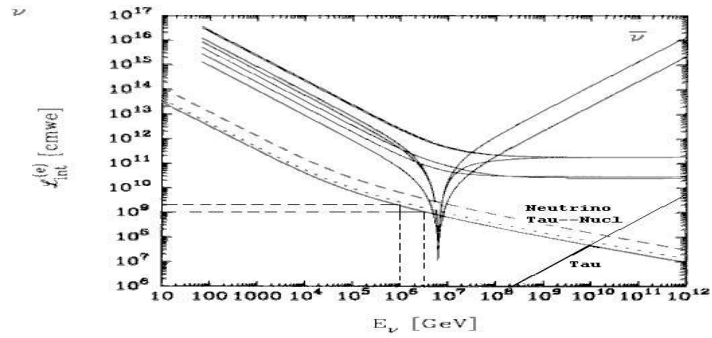}
\end{center}
  \caption {The Gandhi et all (1998) UHE neutrino
    ranges as a function of UHE neutrino energy in Earth with overlapping
    $\bar{\nu}_e e$, $\nu_\tau N$ interactions;
    below the UHE $\tau$ range, as in Fig 1, at the  same energies in matter (water).\newline}
    \label{fig:boxed_graphic 5}
\end{figure}

\begin{figure}[]
\begin{center}
\includegraphics[width=0.8\textwidth,height=4.5cm]{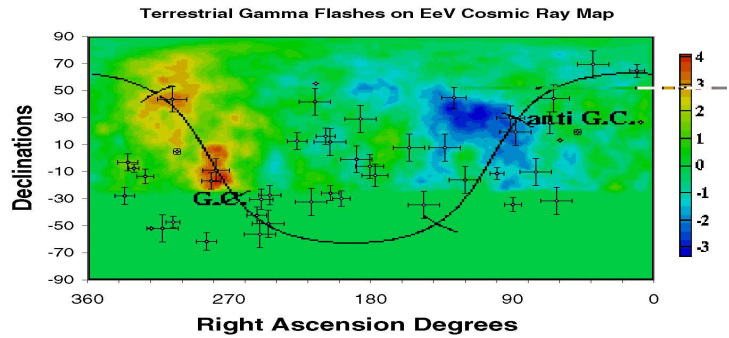}
\end{center}
\caption []{Terrestrial Gamma Flash in celestial coordinate over
UHECR diffused data by AGASA  cosmic rays at EeV energies.}
\label{fig:boxed_graphic 10a}
\end{figure}

\clearpage


\end{document}